\documentclass[journal]{IEEEtran}
\IEEEoverridecommandlockouts
\usepackage{cite}
\usepackage{amsmath,amssymb,amsfonts}
\usepackage{algorithmic}
\usepackage{graphicx}
\usepackage{textcomp}
\usepackage{xcolor}
\usepackage{multirow}
\def\BibTeX{{\rm B\kern-.05em{\sc i\kern-.025em b}\kern-.08em
    T\kern-.1667em\lower.7ex\hbox{E}\kern-.125emX}}
\begin{document}

\title{Cyber-Twin: Digital Twin-boosted Autonomous Attack Detection for  Vehicular Ad-Hoc Networks\\
}

\author{
\IEEEauthorblockN{
Yagmur Yigit \IEEEauthorrefmark{1},
Ioannis Panitsas \IEEEauthorrefmark{2}, 
Leandros Maglaras \IEEEauthorrefmark{1}, 
Leandros Tassiulas\IEEEauthorrefmark{2}, and 
Berk Canberk \IEEEauthorrefmark{1}\\}
\IEEEauthorblockA{
\IEEEauthorrefmark{1}School of Computing, Engineering and The Build Environment, Edinburgh Napier University, United Kingdom \\
 \IEEEauthorrefmark{2} Department of Electrical Engineering, Yale University, New Haven, CT, USA 
 \\
Email: \{yagmur.yigit, L.Maglaras, B.Canberk\}@napier.ac.uk,  \{ioannis.panitsas, leandros.tassiulas\}@yale.edu}
}

\markboth{Accepted by 2024 IEEE International Conference on Communications (ICC), ©2023 IEEE}%
{Shell \MakeLowercase{\textit{et al.}}}

\maketitle

\begin{abstract}
The rapid evolution of Vehicular Ad-hoc NETworks (VANETs) has ushered in a transformative era for intelligent transportation systems (ITS), significantly enhancing road safety and vehicular communication. However, the intricate and dynamic nature of VANETs presents formidable challenges, particularly in vehicle-to-infrastructure (V2I) communications. Roadside Units (RSUs), integral components of VANETs, are increasingly susceptible to cyberattacks, such as jamming and distributed denial of service (DDoS) attacks. These vulnerabilities pose grave risks to road safety, potentially leading to traffic congestion and vehicle malfunctions. 
%
Existing methods face difficulties in detecting dynamic attacks and integrating digital twin technology and artificial intelligence (AI) models to enhance VANET cybersecurity.
%
%
Our study proposes a novel framework that combines digital twin technology with AI to enhance the security of RSUs in VANETs and address this gap. This framework enables real-time monitoring and efficient threat detection while also improving computational efficiency and reducing data transmission delay for increased energy efficiency and hardware durability.
%
%
Our framework outperforms existing solutions in resource management and attack detection. It reduces RSU load and data transmission delay while achieving an optimal balance between resource consumption and high attack detection effectiveness. This highlights our commitment to secure and sustainable vehicular communication systems for smart cities.

\end{abstract}


\section{Introduction}
\label{sec:intro}
In recent years, the advent of Vehicular Ad-hoc NETworks (VANETs) has marked a significant milestone in the evolution of intelligent transportation systems (ITS). These networks, which seamlessly integrate mobile ad hoc networks (MANET), Internet of Things (IoT), and ITS, are at the forefront of revolutionizing road safety and vehicular communication. The critical role of VANETs in reducing traffic accidents and enhancing road safety cannot be overstated, given the alarming statistics from the World Health Organization highlighting the high incidence of road traffic deaths globally \cite{WHO}.

However, the dynamic and complex nature of VANETs introduces significant challenges, particularly in vehicle-to-infrastructure communications \cite{survey2015}. Road Side Units (RSUs), as critical components of VANETs, are often the targets of cyberattacks, including but not limited to jamming and distributed denial of service (DDoS) attacks, which can severely disrupt vehicular communication and pose significant risks to road safety, leading to potential traffic congestion, vehicle malfunctions, and even accidents \cite{22LM}. This vulnerability underscores the urgency for innovative solutions to safeguard these systems against such threats.
The prevalence of connected and autonomous vehicles in modern urban environments underscores the urgent need to address these vulnerabilities due to the rapid evolution of vehicular technology.


\begin{figure*}[t]
    \centering
    \includegraphics[width=7in]{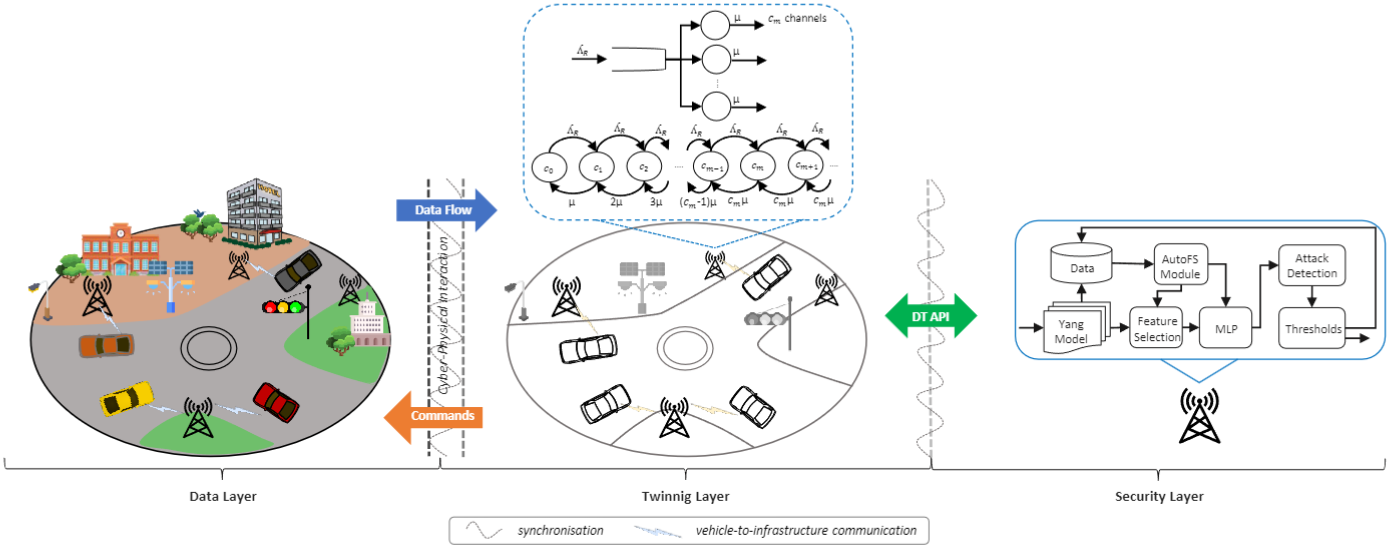}
    \caption{The proposed architecture.}
    \label{fig:system}
\end{figure*}

In addressing these challenges, our work focuses primarily on vehicle-to-infrastructure communication, explicitly targeting the security of RSUs. We introduce a novel framework that employs digital twin technology and an advanced artificial intelligence (AI) model. The digital twin concept, a digital replica of a physical asset or system, is central to our approach \cite{twinPort}. By creating a digital twin of RSUs, we achieve a real-time, dynamic representation of the physical units, which allows for the meticulous monitoring and analysis of network interactions. This capability is crucial in detecting and mitigating the aforementioned cyber threats, ensuring the integrity and reliability of vehicle-to-infrastructure communications \cite{twinpot}. We referred to digital twins as `cyber-twins' since we utilized them in the field of cybersecurity.
%
%
Reducing computational demands and minimizing the average data transmission delay are some of the key areas of research in green VANET technology \cite{greenvanet2023}. Our solution secures communication and contributes to green communications. Our framework focuses on optimizing RSUs' efficiency by reducing their computational demands and minimizing the average data transmission delay. This aligns with the growing emphasis on sustainable and eco-friendly technologies in smart cities. This focus on both security and sustainability places us at the forefront of ITS advancement.
Furthermore, our dedication to creating environmentally friendly technologies for smart cities is highlighted by our efforts to reduce energy consumption and extend the lifespan of hardware.
The key contributions of this paper include:
\begin{itemize}
    \item We propose a novel framework tailored explicitly for securing vehicle-to-infrastructure communication in VANETs, utilizing cyber twins and an AI model for efficient threat detection.
    \item Our approach integrates cyber twins for real-time monitoring, management of vehicular networks, and analysis of RSU performance and security.
    \item Proposed framework reduces RSUs' computational load, improving hardware longevity and energy efficiency.
    \item Our solution significantly advances green communications in VANETs by reducing computational demands, minimizing data transmission delays, and contributing to the sustainability of ITS.
\end{itemize}

Our paper proposes a new approach to enhance vehicle-to-infrastructure communication in VANETs while addressing road safety, sustainability, and security concerns. We use cyber twins and AI models to set a new standard for developing intelligent and secure transportation systems. Our work paves the way for safer and greener smart cities.
This work proceeds with an overview of relevant literature in Section~\ref{sec:related}, then delves into the proposed solution in Section~\ref{sec:proposed}, and evaluates its performance in Section~\ref{sec:performance}. The paper is concluded in Section~\ref{sec:conclusion}.

\section{Related Work}
\label{sec:related}

This section highlights recent advancements in improving road safety and efficiency in VANETs and their integration into ITS, setting the stage for our novel framework.

Alhaidari \emph{et al.} introduced a simulation method to create the VANET DDoS dataset, encompassing critical VANET features like traffic density and node mobility. However, it remains inaccessible to the public \cite{simu21}.
Bangui \emph{et al.} tackled VANETs' vulnerability to malicious jamming, proposing a data prioritization model enhancing Big Data Analytics' efficiency in jamming detection, a leap forward in real-time anti-jamming applications \cite{anti21}.
Mokhtar \emph{et al.} surveyed VANETs' security, highlighting the challenges due to their mobile, infrastructure-less nature \cite{survey2015}. They emphasized the network's inherent security vulnerabilities, paving the way for innovative solutions.
Another notable study introduced a machine learning-based system for real-time identification of malicious nodes in VANETs, achieving high accuracy with random forest and gradient-boosted trees models \cite{sensors2023}.

\begin{figure*}[t]
    \centering
    \includegraphics[width=6.5in]{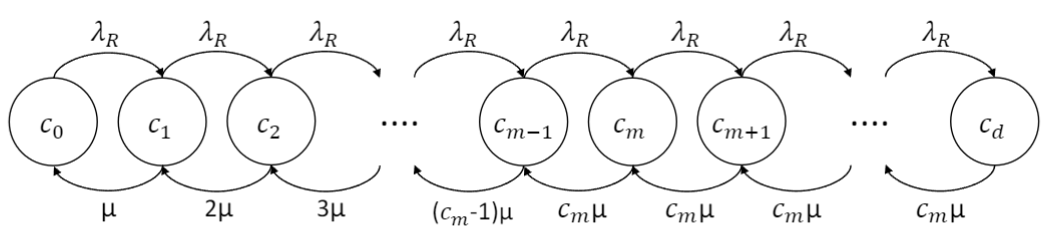}
    \caption{The state diagram.}
    \label{fig:statediagram}
\end{figure*}

Haydari \emph{et al.} developed advanced machine learning methods for intrusion detection in VANETs, targeting sophisticated attacks like DDoS \cite{sensors2022}. Their RSU-based non-parametric detection system has the potential for improvement against sporadic attacks.
Zhou \emph{et al.} concentrated on enhancing VANET security within ITS for 6G systems, introducing a framework combining identity-based encryption and deep learning, primarily addressing DDoS attack scenarios \cite{IEEE2022}. 
Alrehan \emph{et al.} reviewed VANET architectures, advocating for security solutions uniquely tailored to VANET characteristics, diverging from conventional methods \cite{MLSurvey19}. 
Anyanwu \emph{et al.} focused on a software-defined network integrated VANETs' vulnerability to identify DDoS attacks utilizing a support vector machine (SVM) classifier, a significant step in vehicle security \cite{SDN23}. 
A novel self-powered, fog-based VANET infrastructure was also proposed, focusing on sustainable and secure communication, including innovative power management and fog clustering strategies for enhanced network resilience \cite{green23}. 

In VANET research, the utilization of digital twin technology has not yet been widely explored.
Among the few, Arya \emph{et al.} utilized it for identifying malicious nodes, employing machine learning for practical traffic analysis \cite{DT23}. 
Ak \emph{et al.} developed T6CONF, a new framework that enhances IoT communication in smart cities, thereby improving waste management predictions and achieving net-zero waste objectives \cite{T6CONF}.
Additionally, Hu \emph{et al.} explored the internet of vehicles, forming a digital twin connection between physical and virtual vehicles for real-time traffic prediction, highlighting the potential and challenges in handling extensive, sparse data sets \cite{HuIEEEDT22}. 

The studies have highlighted VANETs' immense potential for improving road safety and network security. However, there is a noticeable gap in effectively combining digital twins with AI models to enhance security and environmental sustainability in VANETs. We propose a new framework to secure vehicle-to-infrastructure communication and advance green communication in smart city infrastructures to address this gap.

\section{Proposed System Model for Vehicle-to-Infrastructure Security}
\label{sec:proposed}

Our proposed system is designed to foster a resilient and robust communication framework for vehicles in 6G smart cities, leveraging cyber twins to optimize the computational efficiency of RSUs. This enhances the reliability and stability of the RSUs. Some potential benefits of our system can be:
\begin{itemize}
    \item Enhanced energy efficiency in RSUs, achieved through reduced computational demands and minimizing the average data transmission delay.
    \item Extension of hardware lifespan, resulting in economic benefits due to decreased replacement and maintenance needs.
    \item Greater adaptability of RSUs to emerging technologies, eliminating the need for comprehensive hardware overhauls.
\end{itemize}

Our system adopts a holistic approach with a three-tiered structure comprising physical-to-virtual and virtual-to-virtual communications between layers. This structure forms a comprehensive cyber-twin framework within VANET. Our system is segmented into three layers: data, twinning, and security, each crucial in ensuring the network's functionality and resilience, as seen in Fig.~\ref{fig:system}.

\subsection{Data Layer}
The Data Layer is the foundational segment of our cyber-twin architecture, primarily focusing on data acquisition and transmission. It integrates vehicles and RSUs, collecting critical data for higher-level simulation and security analysis. Its precision in real-time data representation and communication is crucial for maintaining the integrity and effectiveness of the digital twin simulations.


\subsection{Twinning Layer}

This layer is pivotal in creating digital replicas of VANET's physical components. We used queuing theory to model the VANET environment. We have implemented the M/M/m queuing model with a first-in-first-out strategy for efficient data network simulation. The model evaluates the system's state by analyzing communication requests against channel availability, providing insights into average waiting times and queue lengths \cite{BOZKAYA201572}.


In this model, channels are linked to servers, and the system's state is evaluated by comparing the total communication requests against available channels.
For the M/M/m queuing model, our key assumptions and parameters are as follows:
\begin{itemize}
    \item Vehicles share channels equally within a transmission area.
    \item $\lambda_R$ denotes the arrival rate of communication requests.
    \item $\mu$ expresses the channel's service rate per request, which is time-dependent.
    \item $c$ represents the channel states.
    \item $m$ defines the total number of channels.
    \item $d$ denotes the total number of vehicles with communication requests.
\end{itemize}

These parameters feed into the M/M/m model to determine average wait time and queue length metrics. These parameters are crucial in the M/M/m model for assessing the system's efficiency and managing channel requests. 

We formulate basic equations using a state diagram in Fig.~\ref{fig:statediagram}. When the number of vehicle requests $c_d{_i}$ is less than or equal to the number of available channels $c_m$ (i.e., $c_d{_i}  \leq c_m$), all requests are accommodated immediately. However, if the requests exceed the available channels (i.e., $c_d > c_m$), the system processes requests up to the limit of $c_m$, and the remaining requests are queued for subsequent processing. There are $c_m$ channels so the probability is increasing ($\mu, 2\mu, 3\mu \;...\; c_m\mu$) until reaching $c_m$th channel. After the $c_m$th channel, the leaving of the one request stays the same with probability $c_m$th.

The utilization factor for this model can be calculated as follows:
\begin{equation}
\rho = {1 - \frac{\lambda_R}{c_m\mu}}
\label{uti}
\end{equation}

For zero request state:
\begin{equation}
P_0 = \Biggl[ \sum_{c_d=0}^{c_m-1}\frac{(c_m\rho)^{c_d}}{c_d\ !} + \frac{(c_m\rho)^{c_m}}{c_m\ ! (1-\rho)}\Biggr]
\label{p0}
\end{equation}

$P_{c_d}$ is the probability of all channels busy in the system:
\begin{multline}
P_{c_d} =  \Biggl\{ P_0\frac{(c_m\rho)^{c_d}}{c_d\ !} , \qquad c_d \leq c_m 
                    \\
                 P_0\frac{(c_m)^{c_m}(\rho)^{c_d}}{c_m\ !} , \qquad c_d > c_m 
\label{pn}
\end{multline}

The probability of arriving request has to wait in the queue ( $c_m$ request or more in the system):
\begin{equation}
P_{Q_R} = \sum_{c_d=c_m}^{\infty}P_{c_d} = P_0\frac{(c_m\rho)^{c_m}}{c_m\ ! (1-\rho)}
\label{pq}
\end{equation}

When the channel is the full state, the average number of requests in the queue:
\begin{equation}
N_{Q_R} = \frac{P_{Q_R}\rho}{(1-\rho)}
\label{Nq}
\end{equation}

The amount of time spent in the queue, which is the average waiting time in the queue:
\begin{equation}
T_{Q_R} = \frac{P_{Q_R}\rho}{ \lambda_R(1-\rho)}
\label{Tq}
\end{equation}

The total time in the system, which is the sum of queue time and the service time:
\begin{equation}
T = T_{Q_R} + \frac{1}{\mu}
\label{T}
\end{equation}

Lastly, we calculated the average number of channel requests in the system as follows:
\begin{equation}
N_R = {c_m}\rho + \frac{P_{Q_R}\rho}{1-\rho}
\label{Nr}
\end{equation}

Data from the twinning layer is transferred to the security layer via a YANG model for robust data transfer. We used the YANG model from previous work \cite{YADA} and adapted it to VANET.

\subsection{Security Layer}
The system's security mechanism is based on our previous research \cite{YY22}. It includes an AutoFS Module and a Multilayer Perceptron, both essential for effective attack detection.

\subsubsection{AutoFS Module}
This module dynamically selects the most effective Feature Selection (FS) methods under varying network conditions. The system incorporates various techniques such as Recursive Feature Elimination, Backward Feature Elimination, Chi-square,  Fisher Score, and ANOVA F-value Selection, adapting to the dynamic nature of network data. Each FS method has a unique approach to feature selection, influencing DDoS detection efficiency. The module identifies the optimal MLP model and FS method based on system performance metrics.

\subsubsection{Labelling Algorithm}
Our labelling method efficiently categorizes data using the expectation-maximization and K-means algorithms, improving precision. The method clusters and labels data to accurately predict attacks, considering their infrequency.

\subsubsection{Multilayer Perceptron}
Our MLP network adapts to continuously changing network data through online learning. It uses a five-layer architecture with relu and softmax activation functions for effective data classification. It constantly updates model weights, ensuring stable performance in a dynamic network environment.

The proposed system model is an adaptive approach to vehicular communication that bolsters security and optimizes performance for smart city infrastructure. It is a pioneering approach to green and secure 6G smart cities.


\section{Performance Evaluation}
\label{sec:performance}
To test the performance of our proposed system, we used OMNeT++ version 5.1, INET version 3.6, Veins version 4.7, and SUMO version 0.30.0 as the previous work \cite{simu21}. We utilized Eclipse Ditto, which is a scalable and flexible open-source framework, to create cyber twins of physical entities \cite{Eclipse}.
Our evaluation focused on two main aspects: the efficiency of cyber twins in RSU resource utilization and the effectiveness of our framework in attack detection.

\begin{figure}[htbp]
    \centering
    \includegraphics[width=3.45in]{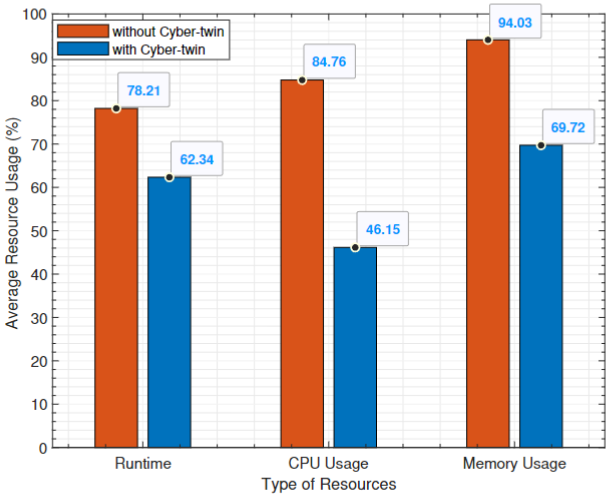}
    \caption{The average resource usage comparison.}
    \label{fig:resource}
\end{figure}

Firstly, we tested the performance of cyber twins in RSU. Therefore, we check the average resource usage of the system with and without the cyber twin. As seen in Fig.~\ref{fig:resource}, when we employed the cyber-twin in the system, the system utilised the resource more efficiently since we reduced the load of the roadside unit. Integrating the cyber-twin system model resulted in more efficient resource utilization, thereby lightening the RSU's load. 

After testing the cyber twin effect of the roadside unit resource performance, we experiment with the attack detection performance of the cyber-twin framework. To this end, we used two datasets. The first dataset is the RF Jamming Dataset \cite{RF-Jamming}, which comprises diverse scenarios of RF jamming attacks and interference in VANET. This dataset includes two datasets for two maximum estimated relative speed types. We combined them, adding a new column with the maximum estimated relative speed.
The ToN-IoT dataset \cite{dataset2}, specifically designed for assessing the reliability and performance of various AI-driven cybersecurity applications, is utilized in our study since Gad \textit{et al.} used this dataset for attack detection in VANET \cite{VANET-ToN}. This dataset is particularly tailored for next-generation IoT and industrial IoT scenarios. However, the current version of the ToN-IoT dataset is not VANET-oriented; therefore, we combine the first dataset's no-attack samples and the ToN-IoT dataset's attack samples. We selectively used a specified number of samples from these datasets to construct Dataset 2. The details of this selection, including the number of samples from each dataset, are meticulously outlined in Table~\ref{tab:dataset}. 

\begin{table}[htbp]
\caption{The Sample Distribution in Dataset 2 
\label{tab:dataset}}
\centering
    \begin{tabular}{|l|c|}
    \hline 
    Dataset Name (Feature) & Number of Samples \\ \hline\hline 
    \begin{tabular}[c]{@{}c@{}}RF Jamming Dataset-1 \\ (No Attack Samples)\end{tabular} & 1000   \\  \hline
    \begin{tabular}[c]{@{}c@{}}RF Jamming Dataset-2\\ (No Attack Samples)\end{tabular}  & 1000   \\  \hline
    \begin{tabular}[c]{@{}c@{}}ToN-IoT Network Dataset\\ (Attack Samples)\end{tabular}  & 400   \\     \hline    
    \end{tabular}
\end{table}

We compare our solution (PS) with KNN and SVM algorithms since they gave the best result in the following works: \cite{Rf-Jam21} and \cite{SVM23}, respectively. Table~\ref{tab:datasetperformance} compares the performance of different methods on the datasets. Our solution outperforms the others.

\begin{table}[htbp]
\caption{The performance comparison of methods 
\label{tab:datasetperformance}}
\centering
    \begin{tabular}{ll|l|l|l|}
 \cline{3-5}
                                                 &     & \begin{tabular}[c]{@{}c@{}}Precision \\ (\%)\end{tabular} & \begin{tabular}[c]{@{}c@{}}F-Measure\\ (\%)\end{tabular} & \begin{tabular}[c]{@{}c@{}}Sensitivity\\ (\%)\end{tabular} \\ \cline{3-5} \cline{3-5} \hline 
    \multicolumn{1}{|l|}{\multirow{3}{*}{Dataset-1}} & \multicolumn{1}{l|}{KNN} & \multicolumn{1}{l|}{95.72}                                & \multicolumn{1}{l|}{96.28}                               & \multicolumn{1}{l|}{96.85}                                 \\ \cline{2-5} 
    \multicolumn{1}{|l|}{}                           & \multicolumn{1}{l|}{SVM} & \multicolumn{1}{l|}{89.27}                                & \multicolumn{1}{l|}{91.84}                               & \multicolumn{1}{l|}{94.58}                                 \\ \cline{2-5} 
    \multicolumn{1}{|l|}{}                           & \multicolumn{1}{l|}{PS}  & \multicolumn{1}{l|}{98.96}                                & \multicolumn{1}{l|}{98.99}                               & \multicolumn{1}{l|}{99.03}                                 \\ \hline \hline
    \multicolumn{1}{|l|}{\multirow{3}{*}{Dataset-2}} & \multicolumn{1}{l|}{KNN} & \multicolumn{1}{l|}{88.67}                                & \multicolumn{1}{l|}{88.90}                               & \multicolumn{1}{l|}{89.14}                                 \\ \cline{2-5} 
    \multicolumn{1}{|l|}{}                           & \multicolumn{1}{l|}{SVM} & \multicolumn{1}{l|}{84.67}                                & \multicolumn{1}{l|}{85.91}                               & \multicolumn{1}{l|}{87.19}                                 \\ \cline{2-5} 
    \multicolumn{1}{|l|}{}                           & \multicolumn{1}{l|}{PS}  & \multicolumn{1}{l|}{97.53}                                & \multicolumn{1}{l|}{98.08}                               & \multicolumn{1}{l|}{98.64}                                 \\ \hline
    \end{tabular}
\end{table}


After that, we examined the performance of our solution in terms of average delay and delivery rate under various data message times.

\begin{figure}[htbp]
   \centering
    \includegraphics[width=3.45in]{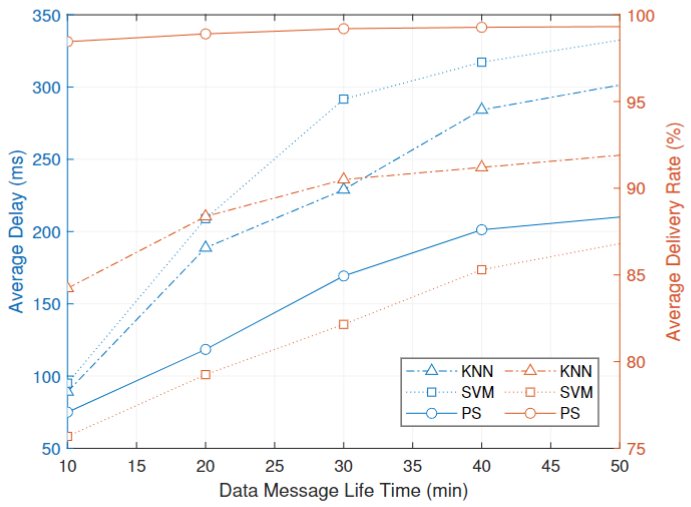}
    \caption{The comparison of methods regarding the average delay and delivery rate under different data message lifetime.}
    \label{fig:delay}
\end{figure}

Fig.~\ref{fig:delay} indicates that PS is performing significantly better in reducing delay than KNN, approximately 1.41 times, and approximately 1.61 times faster than SVM. In terms of average delivery rate, our solution delivers a stable performance. It is approximately 1.11 times more effective than KNN and about 1.21 times more effective than SVM. As can be seen in Fig.~\ref{fig:delay}, PS has significantly lower delays and higher delivery rates compared to the other two methods.

We also investigate the twinning rate performance according to the system's detection rate and total RAM usage. To this end, we used the dataset 2. We define the twinning rate as follows:
\begin{equation}
    \gamma = \frac{\tau}{\sigma}*100
    \label{twinningrate-formula}
\end{equation}
where $\gamma$ is the twining rate, $\sigma$ is the total number of packages in the roadside units in the data layer, and $\tau$ is the total number of packages taken to the twinning layer from the data layer.

\begin{figure}[htbp]
    \centering
    \includegraphics[width=3.45in]{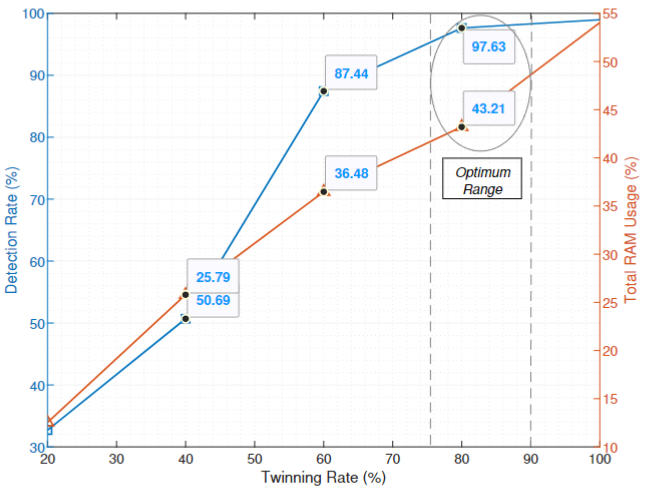}
    \caption{The detection rate and total RAM usage comparison according to the twinning rate.}
    \label{fig:twinning}
\end{figure}

We explored the relationship between the twinning rate, detection rate, and total RAM usage. Fig.~\ref{fig:twinning} depicts the detection rate and total RAM usage comparison according to the twinning rate. According to the results, we defined the optimum twinning rate range between seventy-six and ninety per cent to minimize the total RAM usage while protecting a high attack detection rate. Thus, we established an optimal range that balances resource consumption with high attack detection efficiency by fine-tuning the twinning rate.

In addition to its robust security and efficiency features, our proposed system significantly contributes to green communication practices in 6G smart city infrastructures. The system's advanced cyber twins optimise RSU performance and reduce energy consumption. Moreover, it contributes to green communications by reducing their computational demands and minimizing the average data transmission delay. 
By efficiently managing data traffic and processing demands, the system minimizes the environmental impact of urban communication networks. This reduction in energy requirements aligns with eco-friendly objectives and diminishes operational costs, making it a sustainable solution for future smart cities. Furthermore, the extended lifespan of hardware components due to reduced strain translates into fewer replacements and maintenance needs, further reinforcing the system's environmentally conscious design. The cumulative effect of these factors underscores our system's role in advancing green communication technology, a critical consideration in the development of sustainable urban environments.


\section{Conclusion}
\label{sec:conclusion}

Our research has successfully introduced a novel framework that significantly enhances the security of RSUs in VANETs. We have developed a robust real-time vehicular network framework that integrates cyber twins with advanced AI models. The framework adeptly addresses the complexities of VANETs, particularly in vehicle-to-infrastructure communications. Our research findings demonstrate substantial resource utilization and attack detection improvements, which outperform existing solutions. 
%
The framework's cyber twins of RSUs help detect and mitigate cyber threats, ensuring communication integrity and reliability.
Our framework also contributes to green communications by improving the computational efficiency of RSUs, reducing the average delay of data transmission, and leading to increased energy efficiency and extended hardware durability. It has also achieved an optimal balance between resource consumption and high attack detection effectiveness, with a defined twinning rate range of seventy-six to ninety per cent. These advancements underscore our commitment to developing sustainable, secure, and resilient vehicular communication systems for the future of smart cities. Overall, our framework sets a new benchmark in developing secure and green ITS, addressing security challenges while paving the way for a future of safer and greener smart cities.

\section*{Acknowledgment}
This work is supported by The Scientific and Technological Research Council of Turkey (TUBITAK) 1515 Frontier R\&D Laboratories Support Program for BTS Advanced AI Hub: BTS Autonomous Networks and Data Innovation Lab Project 5239903. 

\bibliographystyle{IEEEtran}


\end{document}